\documentclass[tradiabstract]{aa}            

\usepackage{natbib}
\bibpunct{(}{)}{;}{a}{}{,}
\usepackage{graphicx}
\usepackage{txfonts}

\begin{document}

\title{The architecture of the GJ\,876 planetary system}
\subtitle{Masses and orbital coplanarity for planets b and c
\thanks{Based on observations made with the NASA/ESA \textit{Hubble Space Telescope}, obtained at the Space Telescope Science Institute, which is operated by the Association of Universities for Research in Astronomy, Inc., under NASA contract NAS 5-26555 (programs GO-8102, 8775, and 9233); and on observations obtained at the W.~M.~Keck Observatory, which is operated jointly by the University of California and the California Institute of Technology.}}

\author{J.~L. Bean\inst{1} \and A. Seifahrt\inst{1}}

\institute{Institut f\"{u}r Astrophysik, Georg-August-Universit\"{a}t G\"{o}ttingen, Friedrich-Hund-Platz 1, 37077 G\"{o}ttingen, Germany\\
\email{bean@astro.physik.uni-goettingen.de}}

\date{Accepted January 12, 2009}

\abstract{We present a combined analysis of previously published high-precision radial velocities and astrometry for the GJ\,876 planetary system using a self-consistent model that accounts for the planet-planet interactions. Assuming the three planets so far identified in the system are coplanar, we find that including the astrometry in the analysis does not result in a best-fit inclination significantly different than that found by Rivera and collaborators from analyzing the radial velocities alone. In this unique case, the planet-planet interactions are of such significance that the radial velocity data set is more sensitive to the inclination of the system through the dependence of the interactions on the true masses of the two gas giant planets in the system (planets b and c). The astrometry does allow determination of the absolute orbital inclination (i.e. distinguishing between $i$ and $180-i$) and longitude of the ascending node for planet b, which allows us to quantify the mutual inclination angle between its orbit and planet c's orbit when combined with the dynamical considerations. We find that the planets have a mutual inclination $\Phi_{bc} = 5.0\degr^{\,+3.9\degr}_{\,-2.3\degr}$. This result constitutes the first determination of the degree of coplanarity in an exoplanetary system around a normal star. That we find the two planets' orbits are nearly coplanar, like the orbits of the Solar System planets, indicates that the planets likely formed in a circumstellar disk, and that their subsequent dynamical evolution into a 2:1 mean motion resonance only led to excitation of a small mutual inclination. This investigation demonstrates how the degree of coplanarity for other exoplanetary systems could also be established using data obtained from existing facilities.}


\maketitle

\section{Introduction}
A planetary system's ``architecture'' consists of the census, masses, and orbits of the objects in the system. The formation and evolutionary history of a planetary system are encoded in these characteristics. The architecture of the Solar System is broadly consistent with the theory of planet formation in a circumstellar disk, and its deviations from the expected pattern are interpreted as evidence of evolutionary processes. It is important to determine the architecture of exoplanetary systems so as to have broader constraints on the formation and evolution of planetary systems than can be obtained from study of the Solar System alone.

The GJ\,876 planetary system is one exoplanetary system that has been the focus of much attention to ascertain its architecture. GJ\,876 itself is an M4 dwarf star with an essentially solar metallicity \citep{bonfils05, bean06} that is at least older than 1\,Gyr as evidenced by its low activity and slow rotation \citep{marcy98}. It is at a distance of 4.7\,pc \citep{perryman97} and appears to be a singleton as no stellar companions have been reported. The star does not appear to also harbor a substantial debris disk \citep{trilling00, shankland08}.

A gas giant planet was found to orbit GJ\,876 by two groups independently in 1998 using Doppler spectroscopy \citep{marcy98,delfosse98}. This was the first convincing detection of a planet around an M dwarf, and GJ\,876 still remains one of the few known planet hosting stars of this type. Subsequent observations by \citet{marcy01} revealed that the system contains a second, lower-mass gas giant that is in a 2:1 mean motion resonance with the first detected planet. 

Soon after the discovery of the second planet in the GJ\,876 system, \citet{laughlin01} and \citet{rivera01} pointed out that the two identified planets are experiencing mutual interactions on an unprecedentedly short timescale. They found that the resulting perturbations are of such significance that a model based on Keplerian orbits was not accurate enough to match the radial velocity data, and instead direct integration of the equations of motion (i.e. Newtonian orbits) for the three body configuration was needed. Those authors also found that, although they introduced significant additional complexity in the modeling of the observational data, the occurrence of short-timescale interactions could allow inference of the true masses of the planets from radial velocity data alone. This is in contrast to the usual situation where only planets' minimum masses are determinable from radial velocity data owing the orbital inclination degeneracy. The key to the additional insight is that the non-Keplerian perturbations are dependent on the true masses of the interacting bodies. Therefore, if these perturbations could be characterized well enough with radial velocity data then the planet masses could be constrained. \citet{laughlin01} and \citet{rivera01} both concluded that tight limits on the planet masses could not be set given the data available at the time, but that future analysis of the continuing Doppler measurements for GJ\,876 would possibly give better results.

\citet{benedict02} carried out astrometric observations of the GJ\,876 system using the Fine Guidance Sensor (FGS) instrument on the \textit{Hubble Space Telescope} (\textit{HST}) beginning around the time the first planet was announced and continuing for 2.5 years. Analysis of these data revealed a residual perturbation with semimajor axis of 0.3\,$\pm$\,0.1\,mas in phase with the orbit of planet b expected from modeling radial velocity data. This was the first definitive astrometric detection of an exoplanet, and it is still one out of only a few such successful detections. Modeling the astrometry together with the radial velocities available yielded an estimate of the orbital inclination of planet b ($i_{b} = 84\degr \pm 6\degr$), and thus the planet's true mass after assuming a mass for GJ\,876 ($m_{b} = 1.9 \pm 0.3\,M_{Jup}$). In their analysis, \citet{benedict02} used the standard Keplerian rather than Newtonian orbital calculations for modeling the radial velocity and astrometry data. 

Continuing the trend of exoplanet firsts and rarities in the GJ\,876 system, \citet{rivera05} proposed the existence of a third, very low-mass planet ($m_{d}\,\sin\, i_{d} = 5.9 \pm 0.5 M_{\oplus}$) based on a relatively extensive set of new high-precision radial velocities. At the time, this was possibly (allowing for the inclination ambiguity) the lowest mass planet yet found around a main sequence star, and it is still one of only a few known planets with a mass potentially in the ``Super-Earth'' regime (i.e. $1\,\la\,m\,\la\,10\,M_{\oplus}$). A comprehensive photometric search for transits of this planet by the discovery team did not result in a detection, although grazing transits could not be ruled out.

\citet{rivera05} modeled their radial velocity data with self-consistent Newtonian four-body orbits assuming the three identified planets were in coplanar orbits. This modeling of the new data set yielded seemingly tight constraints on the inclination of the system as foreseen by \citet{laughlin01} and \citet{rivera01}. \citet{rivera05} found that the coplanar system inclination appeared to be $\sim50\degr$, and the masses for planets b, c, and d were 2.3\,$M_{Jup}$, 0.8\,$M_{Jup}$, and 7.5\,$M_{\oplus}$ respectively. Although no formal uncertainty in the inclination was given, inspection of their reported $\chi^{2}$ map (see their Fig.\,3) suggests a standard error of $\sim2\degr$.

\citet{rivera05} also argued that the orbits of the two gas giants must have a small or even zero mutual inclination (i.e. they are coplanar) because the radial velocity fit quality of their model deteriorated when the inclinations of each of the planets were pushed away from 50$\degr$. However, quantification of the mutual inclination of the planets' orbits was not possible due to the incomplete characterization of their full three dimensional orbits. Specifically missing was needed information on the planets' absolute orbital inclinations ($i$ or $180-i$) and orbital longitudes of the ascending nodes. \citet{rivera05} did not consider the \textit{HST} astrometry or the results from \citet{benedict02} in their analysis.

The findings of \citet{benedict02} and \citet{rivera05} with regards to the planets' orbital inclinations and masses are inconsistent, and this has led to confusion about what is the ``best'' model of the GJ\,876 system. Numerous theoretical studies have been carried out since the discovery of the second planet to determine what physical processes gave rise to the system's unique architecture \citep[e.g. ][]{lee02, ji02, kley04, kley05, zhou05, crida08} because such specific consideration has the potential for constraining general theories of planet formation and evolution. Determining what mechanisms could have led to the current system arrangement depends critically on knowing the masses of the planets involved, and what the current arrangement itself even is. Therefore, continued refinement of the GJ\,876 system model would be valuable. 

In this paper we present new constraints on the architecture of the GJ\,876 planetary system based on reanalysis of the previously published high precision radial velocities and astrometry for the system. We aimed to resolve the discrepancy between the results of \citet{benedict02} and \citet{rivera05}, and study the degree of coplanarity in the system by the first combined analysis of the data sets using a self-consistent Newtonian orbit model. Our study represents a synergy in the conceptual advances made by \citet{benedict02} in their analysis of a combined radial velocity and astrometry data set for an exoplanetary system, and that of \citet{laughlin01}, \citet{rivera01}, and \citet{rivera05} in their use of Newtonian orbits to account for the signature of multi-body interactions in radial velocity data and infer more information than it is normally possible to obtain from such data. The paper is organized as follows. In \S2 we describe the data utilized in our analysis. We present our analysis in \S3. We conclude in \S4 with a discussion of the implications of the results obtained and the potential for continued work in this area.

\section{The data}
\subsection{Radial velocities}
We utilized the time series radial velocities for GJ\,876 presented by \citet{rivera05} in our analysis. These velocities were measured from high-resolution spectroscopic observations made with the HIRES instrument equipped with an iodine absorption cell and fed by the Keck\,I telescope at the W.~M.~Keck Observatory. This data set contains 155 measurements that have a median uncertainty of 4.1\,m\,s$^{-1}$ and were obtained over 7.6\,yr. No adjustments were made to the uncertainties to account for the potential affect of stellar ``jitter'' (a loose term referring to changes in stellar spectra arising from variation of inhomogeneities on the surface of stars that can be misconstrued as a radial velocity change) because \citet{rivera05} achieved a best-fit reduced $\chi^{2}$ for the radial velocities very close to 1.0, which indicates that there is little or no additional noise in the data. More details about these data can be found in \citet{rivera05} and references therein. 

Other relevant radial velocity data for GJ\,876 were presented by \citet[][data from the ELODIE and CORALIE spectrographs]{delfosse98} and \citet[][data from Lick Observatory]{marcy01}. We elected not to include these data sets in our analysis because their consideration would have been more confusing than illuminating. As all the radial velocities are relative in nature, each data set included therefore requires the addition of a free offset parameter in the orbit fitting. Furthermore, there is the issue of how to treat the estimated uncertainties when using inhomogeneous data sets. Each group has their own method of error estimation based on some combination of the photon statistics in the obtained spectra, stability of the instrument used, and accuracy of the Doppler shift measurement method employed. Also, the uncertainties for the velocities in the three neglected data sets are all $>$ 10\,m\,s$^{-1}$ and none approach the time baseline of the Keck velocities . Therefore, our choice to use only this later data set simplifies the analysis and bypasses a number of potential issues without ignoring very useful data. We note that \citet{rivera05} also chose to focus exclusively on the Keck velocities for GJ\,876 in their analysis. 

\subsection{Astrometry}
We also included in our study the astrometry for GJ\,876 that was obtained with the FGS3 instrument on the \textit{HST} and that was presented by \citet{benedict02}. The data set is made up of observations of GJ\,876 and five reference stars obtained over 27 \textit{HST} orbits and distributed in 9 epochs. The measurements span 2.5\,yr and are coincident with the Keck radial velocities. The majority of the observations were obtained near the periastron, apastron, and subsequent periastron times of planet b during one of its orbits (a time span of $\sim$60\,d). Further observations were scheduled to allow breaking the degeneracy between the orbital, proper, and parallactic motions.

We used the exact same data as \citet{benedict02} with only one modification. We multiplied the nominal uncertainties estimated by their data reduction pipeline for the X and Y axis position measurements by 0.34 and 0.50 respectively. These re-weightings were motivated when we obtained a reduced $\chi^{2}$ significantly below 1.0 for the data during preliminary modeling. The weightings were iteratively adjusted to yield a reduced $\chi^{2} = 1.0$ for the best-fit coplanar model (see below). The separate re-weightings for the X and Y axis data are appropriate because the FGS instrument has two separate arms for measuring the apparent positions in the two axes. Thus, the data from the axes are essentially independent. The median uncertainty in the position measurements for both axes is 0.95\,mas after re-weighting. More details about these data, and FGS measurements in general, can be found in \citet{benedict02} and references therein. 

\section{Analysis}
\subsection{Modeling}
Our analysis consisted of modeling the Keck radial velocities and \textit{HST} astrometry described in \S2 simultaneously. A self-consistent model for a four-body system (three planets and the host star) was used to account for the orbital motion of GJ\,876 in both data sets. This model was generated using the Mercury code \citep{chambers99} to integrate the equations of motion. All the bodies were assumed to be point masses and the only force considered was Newtonian gravity. We have previously used this same general method to simultaneously model radial velocities and eclipse times for an exoplanetary system with consideration of possible planet-planet interactions \citep{bean08}.

We assumed the mass of GJ\,876 ($m_{A}$) is 0.32\,$M_{\sun}$, as suggested by \citet{rivera05} based on consultation with empirical Mass -- Luminosity relationships for low mass stars. The uncertainty in the estimate is probably $\sim$10\%. We did not attempt to account for this uncertainty because it would be prohibitively time consuming to repeat the analyses numerous times with different assumed values. With this assumed mass, the model for the orbital motion of GJ\,876 depended on the input masses and osculating orbital elements for the three planets. The orbital elements are the six usual ones: semi-major axis ($a$), eccentricity ($e$), argument of periastron ($\omega$), mean anomaly ($M$), inclination ($i$), and longitude of the ascending node ($\Omega$). The reference epoch for the osculating elements was taken to be HJD 2\,452\,490.0 as \citet{rivera05} did so that our results may be directly compared to theirs. Including the masses, there were 7 parameters for each planet and 21 parameters total in the orbit model.

Following \citet{rivera05}, we always fixed the eccentricity and argument of periastron for planet d to zero. This planet is expected to be in a nearly circular orbit owing to tidal torques from the host star. Furthermore, it induces a modulation with semi-amplitude of only 6.5\,m\,s$^{-1}$ on the radial velocities. Therefore, the signature of non-zero eccentricity is negligible given the quality of the data and may be ignored. The treatment of the remaining orbit model parameters varied in the different analyses described below. 

Comparison of the orbit model with the radial velocities required one additional step. A single correction factor was added to all the model radial velocities to shift them to the relative scale of the observed velocities. This offset was always a free parameter in the analyses described below. The equation of condition for the radial velocity model was thus
\begin{equation}
\Delta_{\gamma} = RV - (\gamma + ORBIT_{R}),
\label{eq1}
\end{equation}
where $RV$ is the measured relative radial velocities, $ORBIT_{R}$ is the model radial velocity component of GJ\,876's orbital motion, $\gamma$ is the offset, and $\Delta_{\gamma}$ is the residual.

Our approach for generating the astrometric model closely followed the methods used by \citet{benedict02}, which have also been used to analyze FGS astrometry for other exoplanetary systems \citep{mcarthur04, benedict06, bean07} and binary star systems \citep[e.g.][]{benedict01}. The model included the orbital motion of GJ\,876, parallactic and proper motion for GJ\,876 and the five reference stars, and plate adjustments for the 27 epochs of data. 

We selected the observations made during epoch 22 to serve as the astrometric constraint ``plate.'' FGS observations of different stars during an epoch are carried out sequentially rather than simultaneously so there isn't an actual plate in the traditional sense. However, the sequential observations during a single \textit{HST} orbit may be combined to form an effective plate due to stability of the telescope and instrument response during that time period. We refer to this combination of sequential observations as simply a plate below for brevity. The choice of the constraint plate does not have a significant impact on the results and our specific choice was made for consistency with \citet{benedict02}.

The position deviations of the stars due to parallactic, proper, and orbital (GJ\,876 only) motion were calculated in the usual right ascension -- declination reference frame first. These deviations were then rotated about the roll angle of the FGS during the constraint plate observations to place them in the X -- Y reference frame of the instrument at that epoch. The equations used for these calculations were
\begin{equation}
D_{\alpha} = P_{\alpha}\pi + \mu_{\alpha}\Delta t + ORBIT_{\alpha},\label{eq2}
\end{equation}
\begin{equation}
D_{\delta} = P_{\delta}\pi + \mu_{\delta}\Delta t + ORBIT_{\delta},\label{eq3}
\end{equation}
\begin{equation}
D_{\xi} = D_{\alpha}\,\cos\,\theta + D_{\delta}\,\sin\,\theta,
\label{eq4}
\end{equation}
\begin{equation}
D_{\eta} = -D_{\alpha}\,\sin\,\theta + D_{\delta}\,\cos\,\theta,
\label{eq5}
\end{equation}
where $D$ are the motion displacements, $P$ are the parallax factors, $\pi$ is the parallax, $\mu$ are the proper motions, $\Delta t$ is the time difference from the reference epoch, $ORBIT$ are the orbital motions, and $\theta$ is the roll angle of the constraint plate. The $\alpha$ and $\delta$ subscripts refer to the right ascension and declination components respectively. The $\xi$ and $\eta$ subscripts refer to the X and Y components in the reference frame of the constraint plate respectively.

\begin{table*}[t!]
\caption{Parameters from the coplanar analysis.}
\label{t1}
\begin{center}
\begin{tabular}{lccc}
\hline
\hline \\[-3mm]
\textit{Orbital Parameters} \\
Parameter & Planet b & Planet c & Planet d \\
\hline\\[-1mm]
$m$       &  2.57$^{+0.06}_{-0.08}$\,$M_{Jup}$ &  0.80$^{+0.02}_{-0.02}$\,$M_{Jup}$  &  8.17$^{+0.95}_{-0.93}$\,$M_{\oplus}$ \\[2mm]
$a$\, (AU)  &  0.20688$^{+0.00005}_{-0.00004}$ & 0.13062$^{+0.00004}_{-0.00004}$ & 0.0208069$^{+0.0000001}_{-0.0000004}$ \\[2mm]
$e$        & 0.0376$^{+0.0022}_{-0.0019}$ & 0.2657$^{+0.0022}_{-0.0017}$ & 0.0 (Fixed) \\[2mm]
$\omega$\, ($\degr$) & 184.0$^{+2.8}_{-3.3}$ & 197.3$^{+0.4}_{-0.6}$ & 0.0 (Fixed) \\ [2mm]
$M$\, ($\degr$) & 167.3$^{+3.8}_{-3.2}$ & 311.6$^{+1.0}_{-0.8}$ & 311.8$^{+4.4}_{-5.9}$ \\ [2mm]
\hline\\[-3mm]
Parameter & Value &  &  \\
\hline \\[-1mm]
$i$\, ($\degr$) & 48.9$^{+1.8}_{-1.6}$ &  &  \\[2mm]
$\Omega$\, ($\degr$) & 251$^{+16}_{-16}$ &  &  \\[2mm]
\hline
\hline\\[-3mm]
\textit{Fit information} \\
Parameter & Value &  &  \\
\hline\\[-1mm]
$\chi^{2}$   & 860.0 & &  \\[2mm]
DOF          & 843 & & \\[2mm]
radial velocity rms (m\,s$^{-1}$) & 4.2 & & \\[2mm]
astrometry rms (mas) & 0.9 & & \\[2mm]
\hline
\end{tabular}
\end{center}
Note: The orbital parameters are osculating and valid at HJD\,2\,452\,490.0.
\end{table*}

The roll angle of the constraint plate was estimated to be $26.01\degr \pm 0.05\degr$ based on comparison to ground based astrometry catalogs. The actual roll angle we used for calculating the model was always a free parameter and the estimated value was treated as an observation with error to provide a constraint (i.e. a comparison of the estimated value to the actual value used was included in the overall $\chi^{2}$ calculation).

We always solved for the parallaxes and proper motions of GJ\,876 and the five reference stars in each of the analyses described below. We used as observations with error the same estimated parallaxes and previously measured proper motions for the reference stars also utilized by \citet{benedict02}. Unlike \citet{benedict02}, we also used the \textit{Hipparcos} parallax and proper motion for GJ\,876 \citep[$\pi$ = 212.69 $\pm$ 2.10\,mas, $\mu_{\alpha}$ = 960.31 $\pm$ 3.77\,mas\,yr$^{-1}$, $\mu_{\delta}$ = -675.61 $\pm$ 1.58\,mas yr$^{-1}$,][]{perryman97} as observations with error. This is justified because the \textit{Hipparcos} solution for GJ\,876 is unlikely to be affected by its orbital motion due to the small size ($\sim$0.3\,mas) and short timescale ($\sim$60\,d) of the perturbations relative to the precision ($\sim$5\,mas) and time span (2.2\,yr) of the \textit{Hipparcos} observations of GJ\,876.

We used the same six parameter model as \citet{benedict02} to account for changes in the plate scale, rotation, and offset during the different observational epochs. The equations of condition for the astrometry model were
\begin{equation} 
\Delta_{\xi} = Ax + By + C + R_{x}(x^{2} + y^{2}) - (\xi + D_{\xi}),
\label{eq6}
\end{equation}
\begin{equation} 
\Delta_{\eta} = -Bx + Ay + F + R_{y}(x^{2} + y^{2}) - (\eta + D_{\eta}),
\label{eq7}
\end{equation}
where $\Delta$ are the residuals, $x$ and $y$ are the measured positions; $A$, $B$, $C$, $F$, $R_{x}$, and $R_{y}$ are the plate parameters; and $\xi$ and $\eta$ nominal positions of the stars at the reference epoch. For the observations made during the adopted constraint epoch, the plate parameter $A$ was fixed to 1, and $B$, $C$, $F$, $R_{x}$, and $R_{y}$ were fixed to 0. These were free parameters for each of the other 26 epochs. The nominal positions for each star at the reference epoch were also free parameters. 

In total, and excluding the orbital motion component for GJ\,876, there were always five free parameters for each of the six stars, six free parameters for each of 26 epochs, plus the one roll angle for a total of 186 astrometric only parameters. There were 436 FGS observations in each X and Y, as well as three constraints for each of the five stars and one constraint for the roll angle.

We used the usual $\chi^2$ parameter as the goodness-of-fit metric in our analyses. The $\chi^2$ of the model comparisons to all the data was calculated from Eq. (\ref{eq1}), (\ref{eq6}), and (\ref{eq7}) along with the comparison of the certain astrometric parameters mentioned above to their input values.

\subsection{Coplanar study}
We carried out an analysis of the data assuming that the planets were in coplanar orbits (i.e. we set $i = i_{b} = i_{c} = i_{d}$ and $\Omega = \Omega_{b} = \Omega_{c} = \Omega_{d}$). We used a combination of grid search and local minimization algorithms to find the parameters that minimized the $\chi^2$ between the model and the observed data. The parameter uncertainties were estimated by stepping out from the best-fit values of each parameter in turn while marginalizing over the remaining parameters until the $\chi^2$ increased by 1.0 from the minimum value. The identified orbital parameters and their uncertainties along with some fit quality statistics are given in Table~\ref{t1}. The identified astrometric parameters (parallaxes, proper motions, and positions) are essentially the same as for the non-coplanar analysis (\S3.3), so we give only the results from the later investigation because we consider them more robust (see Table~\ref{t2}).

For the orbital orientations, we find $i = 48.9\degr^{\,+1.8\degr}_{\,-1.6\degr}$ and $\Omega = 251\degr^{\,+16\degr}_{\,-16\degr}$. The inclination we determine for a coplanar model is very similar to that suggested by \citet{rivera05} based on analysis of the radial velocities alone. It is also completely consistent with the astrometric perturbation size that \citet{benedict02} measured, but is not consistent with their estimated inclination for planet b ($i_{b} = 84\degr \pm 6\degr$). The difference between our result and theirs arises from the radial velocity data. We have more and better quality data than was available then. In addition, we have the benefit of hindsight that the dynamical modeling is crucial for the radial velocity data, and that this modeling gives more constraints on the planets' orbital orientations than can usually be obtained.

\begin{figure}
\resizebox{\hsize}{!}{\includegraphics{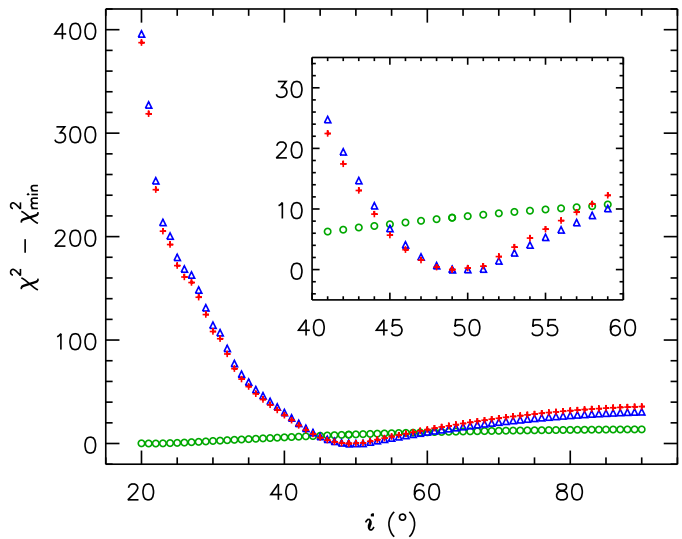}}
\caption{Relative change in the best-fit total $\chi^{2}$ (pluses), radial velocity component $\chi^{2}$ (triangles), and astrometry component $\chi^{2}$ (circles) as a function of the coplanar model inclination. The inset shows a region around the minimum for the total $\chi^{2}$.}
\label{f1}
\end{figure}

We find that the planet-planet perturbations are of such significance that the radial velocity data set is actually more sensitive than the astrometry to the inclination of the system in this unique case. This is due to the dependence of the interactions, which are visible in the radial velocity data, on the true masses of the two gas giant planets (planets b and c). The situation is illustrated in Fig.~\ref{f1}, where we show the best-fit $\chi^2$ for the radial velocity and astrometry data components along with the total for fixed inclinations.  The perturbation due to planet b is clearly detected in the astrometry (with false alarm probability of 1.5 x 10$^{-5}$), but its inclusion does not alter the fit quality response so much that the best-fit inclination is significantly different because of the steep response of the radial velocity fit quality.

The astrometry fit quality component does trend lower for smaller inclinations, but this should not be interpreted to mean that the astrometry would favor smaller inclinations were it independent of the radial velocities. This is because the astrometry data are not sufficient to uniquely determine all of the necessary planet orbital parameters. Therefore, they cannot be divorced from the constraints given by radial velocities and an ``astrometric only'' inclination cannot be determined. For example, the astrometry data are better fitted for $i=30\degr$, but only with the majority of the orbital parameters determined from the fit to the radial velocities. In the later case, the fit is very bad ($\Delta\chi^{2} > 100$ from the best-fit) and, thus, the orbital parameters determined for this inclination and used to generate the astrometric orbit are not physical. As the fit to the astrometry is still acceptable for $i\,\sim\,50\degr$ \citep[we have only a 0.05\,mas larger rms than the best-fit of ][]{benedict02}, we conclude that the astrometry and radial velocities are not in disagreement and that the model giving the best-fit to the combined data set is the optimal one. 

Although the radial velocities are sensitive to the inclination of the system, there remains a degeneracy between $i$ and $180\degr - i$ pairs that is unresolvable with that data alone. This is because the radial velocities are sensitive to the inclination through their dependence on the true masses of the planets and the masses would be the same for $i$ and $180\degr - i$. The inclusion of the astrometry resolves this degeneracy, and we find that the inclination of the system is near 50\degr rather than 130\degr. 

\subsection{Constraining the degree of coplanarity}
\subsubsection{Fitting the data}
As discussed in \S1, \citet{rivera05} found that the radial velocities are sensitive to differences in the orbital inclinations of planets b and c, and they reached a tentative conclusion that the orbits were likely coplanar. However, the mutual inclination angle ($\Phi$) of two orbits depends not only on the inclinations, but also on the longitudes of the ascending nodes. In this case, the mutual inclination of planets b's and c's orbits is given by the equation
\begin{equation} 
\cos\,\Phi_{bc} = \cos\,i_{b}\,\cos\,i_{c} + \sin\,i_{b}\,\sin\,i_{c}\,\cos\,(\Omega_{b} - \Omega_{c}).
\label{eq8}
\end{equation}
Because of their lack of the constraint on the absolute orbital inclinations and orbital longitudes of the ascending nodes that is offered by the astrometry, \citet{rivera05} were unable to quantify the mutual inclination of the two planets' orbits.

In our case, the inclusion of the astrometry helps characterize the full three dimensional orbit of planet b. Therefore, we have a benchmark to measure misalignment relative to. This motivated us to use the combined data set and dynamical model to determine the orbital orientations of planets b and c uniquely, and thus set limits on their orbital mutual inclination.

To do this, we fit the data using an expanded version of the coplanar model. Instead of solving for a single inclination and longitude of the ascending node, we allowed planets b and c to have their own independent values ($i_{b}$, $\Omega_{b}$, $i_{c}$, and $\Omega_{c}$). The data are not very sensitive to the orientation of planet d's orbit, so we set its inclination and longitude of the ascending node to that of planet b (i.e. $i_{d}$ = $i_{b}$ and  $\Omega_{d}$ = $\Omega_{b}$), which is the most massive of the three planets. 

For our initial analysis, we used the same grid search and local minimization technique as for the coplanar study. However, we found the $\chi^2$ surface to be very chaotic. This hindered reliable identification of the minimum $\chi^2$ and error estimation because there were many valleys in the surface containing local minima that were statistically indistinguishable from each other.

The difficulty with the previous method led us to ultimately use a Markov Chain Monte Carlo (MCMC) analysis to identify the most likely parameter values and their confidence intervals for the non-coplanar model. The details and advantages of MCMC are described extensively elsewhere \citep[for discussion in an astronomical context see e.g. ][]{tegmark04,ford05}. Our implementation used 10 Markov chains of 10$^{7}$ points each. For each chain step, a jump in one of the parameters was considered. If the jump resulted in a lower $\chi^2$ than the previous point the jump was accepted. Otherwise, the jump was accepted with a probability of $\exp(-\Delta\chi^2/2)$. The characteristic jump sizes for each parameter were tuned to give a 20 -- 40\% acceptance rate.

\begin{table*}[t!]
\caption{Parameters from the non-coplanar analysis.}
\label{t2}
\begin{center}
\begin{tabular}{lccccc}
\hline
\hline
\textit{Orbital Parameters} \\
Parameter & Planet b & Planet c & Planet d & & \\
\hline\\[-1mm]
$m$       &  2.64$^{+0.11}_{-0.09}$\,$M_{Jup}$ &  0.78$^{+0.05}_{-0.03}$\,$M_{Jup}$  &  8.41$^{+0.78}_{-0.75}$\,$M_{\oplus}$ & & \\[2mm]
$a$\, (AU)  &  0.20700$^{+0.00010}_{-0.00009}$ & 0.13062$^{+0.00005}_{-0.00005}$ & 0.0208069$^{+0.0000004}_{-0.0000004}$ & &  \\[2mm]
$e$        & 0.0363$^{+0.0028}_{-0.0026}$ & 0.2683$^{+0.0058}_{-0.0052}$ & 0.0 (Fixed)  & & \\[2mm]
$\omega$\, ($\degr$) & 188.2$^{+4.9}_{-4.0}$ & 200.4$^{+1.8}_{-1.9}$ & 0.0 (Fixed)  & & \\ [2mm]
$M$\, ($\degr$) & 163.1$^{+4.0}_{-4.9}$ & 309.1$^{+1.9}_{-1.7}$ & 312.2$^{+4.9}_{-5.0}$  & & \\ [2mm]
$i$\, ($\degr$) & 47.2$^{+2.2}_{-2.4}$ & 51.1$^{+3.6}_{-3.9}$ & 47.2 (Tied)  & & \\[2mm]
$\Omega$\, ($\degr$) & 252.3$^{+8.4}_{-7.7}$ & 249.4$^{+7.1}_{-8.4}$ & 252.3 (Tied)  & & \\[2mm]
\hline
\hline
\textit{Astrometric Parameters}\\
Star   &  $\xi$   &   $\eta$   &  $\pi_{abs}$  &   $\mu_{\alpha}$   &  $\mu_{\delta}$ \\
      &  (arcsec)    &   (arcsec)   &  (mas)  &   (mas\,yr$^{-1}$)   &  (mas\,yr$^{-1}$) \\
\hline\\[-1mm]
GJ\,876 & 51.9001$^{+0.0004}_{-0.0004}$ & 730.3929$^{+0.0003}_{-0.0003}$ & 215.5$^{+0.4}_{-0.5}$ & 955.7$^{+1.7}_{-1.7}$ & -673.4$^{+1.1}_{-1.1}$ \\[2mm]
Ref-2 & -27.8463$^{+0.0003}_{-0.0003}$ & 767.6817$^{+0.0005}_{-0.0005}$ & 1.0$^{+0.3}_{-0.3}$ & 5.5$^{+1.6}_{-1.7}$ & -16.0$^{+1.2}_{-1.1}$ \\[2mm]
Ref-3 & -226.2855$^{+0.0005}_{-0.0005}$ & 759.8752$^{+0.0007}_{-0.0007}$ & 3.1$^{+0.4}_{-0.4}$ & 14.2$^{+2.0}_{-2.0}$ & 2.4$^{+1.6}_{-1.7}$ \\[2mm]
Ref-4 & -297.4706$^{+0.0005}_{-0.0005}$ & 639.3754$^{+0.0005}_{-0.0005}$ & 2.3$^{+0.6}_{-0.6}$ & -39.8$^{+2.2}_{-2.2}$ & -43.0$^{+1.2}_{-1.2}$ \\[2mm]
Ref-5 & 450.9640$^{+0.0006}_{-0.0006}$ & 635.2999$^{+0.0004}_{-0.0004}$ & 4.7$^{+0.6}_{-0.7}$ & 7.3$^{+3.0}_{-3.3}$ & -1.2$^{+2.4}_{-2.2}$ \\[2mm]
Ref-6 & 351.6679$^{+0.0005}_{-0.0005}$ & 594.2625$^{+0.0005}_{-0.0005}$ & 1.9$^{+0.3}_{-0.3}$ & -13.1$^{+2.4}_{-2.7}$ & -3.8$^{+2.3}_{-2.1}$ \\[2mm]
\hline
\hline
\textit{Fit information} \\
Parameter  &  Value & & & & \\
\hline\\[-1mm]
$\chi^{2}$   & 855.8  & & & & \\[2mm]
DOF          & 841    & & & & \\[2mm]
radial velocity rms (m\,s$^{-1}$) & 4.2  & & & &\\[2mm]
astrometry rms (mas) & 0.9  & & & &\\[2mm]
\hline
\end{tabular}\\[1mm]
\end{center}
Note: Planet d's inclination and longitude of the ascending node were tied to those of planet b. The orbital parameters are osculating and valid at HJD\,2\,452\,490.0. The coordinates ($\xi$ and $\eta$) are relative positions in the reference frame of the constraint plate. To convert to the RA -- DEC reference frame, the coordinates should be rotated about the inverse of the determined roll angle ($26.07\degr^{\,+0.04\degr}_{\,-0.04\degr}$). The $\chi^{2}$ and rms shown are for the best-fit model.
\end{table*}

Each chain was initialized with a different combination of parameter values well dispersed from the region of parameter space thought to contain the lowest $\chi^2$. Initial tests indicated a typical correlation length of $\sim$2000 points, or roughly 10 times the number of parameters. Therefore, we elected to record the parameters only every 2000 steps to save memory. Each chain took 30 CPU days computation time on an average desktop computer. 

All the chains converged to the same region of parameter space, or ``burned in'', within $\sim$18\,000 steps. To provide a statistical check that the probability distributions had been thoroughly sampled, we computed the \citet{gelman92} $R$ statistic for the parameter values among the chains. The statistic was within 10\% of unity for all the parameters, which indicates the chains were likely long enough for robust inference. 

After trimming the burn-in points, we combined the data from the 10 chains to give parameter distributions with 49\,910 points. We adopted the medians of the MCMC distributions as the best estimates of the parameter values. The 1$\sigma$ uncertainties were taken to be the range of values that encompassed 68.3\% of the parameter distributions on each side of the corresponding median. The results are given in Table~\ref{t2}. It should be noted that the errors from the MCMC analysis are \textit{correlated} because they are calculated from the distributions arising from a simultaneous determination of all the parameters. This explains why the errors on our astrometric parameters are larger by factors of 2 -- 3 than the \textit{uncorrelated} errors given by \citet{benedict02} despite our achieving a similar astrometric fit quality (rms = 0.9\,mas).

Further support for the validity of the MCMC analysis is the fact that the orbital model formed by the adopted parameter values (medians of the MCMC parameter distributions) is essentially the same as the one we initially identified as the best non-coplanar model using the grid search with local minimization technique ($\Delta\chi^{2}$ = 0.2). The main advantages of the MCMC method are that the parameter confidence limits account for the irregular $\chi^2$ surface, and allow calculation of composite parameter uncertainties when including correlations among the parameters (see below).

\begin{figure*}[t!]
\resizebox{\hsize}{!}{\includegraphics{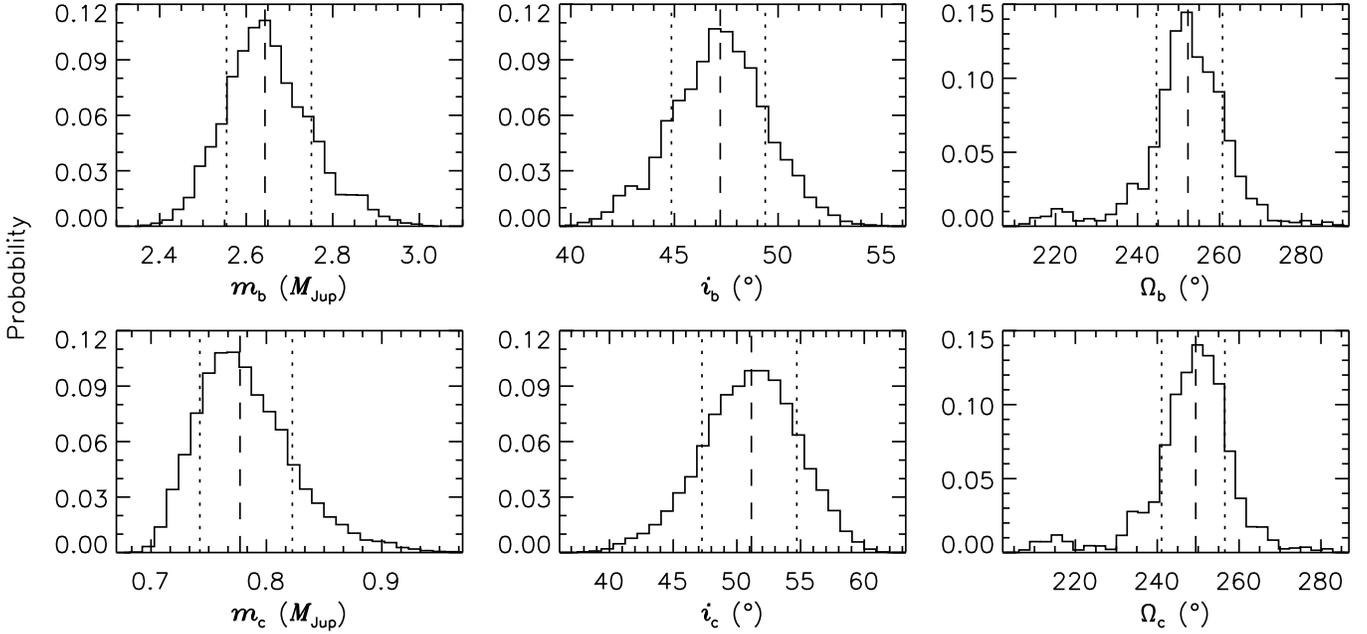}}
\caption{Probability distributions for the masses, inclinations, and longitudes of the ascending nodes for planets b and c from the MCMC analysis. The medians and two-sided 68.3\% confidence limits are given by the dashed and dotted lines respectively.}
\label{f2}
\end{figure*}

\begin{figure}[ht!]
\resizebox{\hsize}{!}{\includegraphics{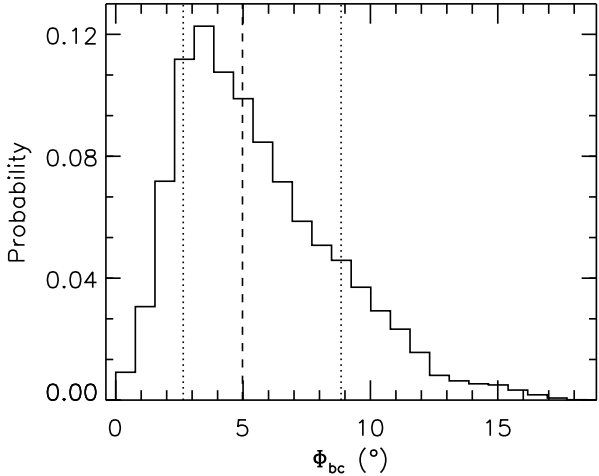}}
\caption{Probability distribution for the mutual inclination of the orbits of planets b and c computed from the probability distributions for $i_{b}$, $\Omega_{b}$, $i_{c}$, and $\Omega_{c}$ using Eq.~\ref{eq8}. The median and two-sided 68.3\% confidence limits are given by the dashed and dotted lines respectively.}
\label{f3}
\end{figure}

The MCMC parameter distributions for the masses, inclinations, and longitudes of the ascending nodes for planets b and c are shown in Fig.~\ref{f2}. As these plots demonstrate, the combined analysis of the radial velocity and astrometry data with the dynamical model yielded tight constraints on the masses and orbital orientations of the two planets. Using these probability distributions, we may calculate the probability distribution for the mutual inclination angle between their orbits directly. The result is shown in Fig.~\ref{f3}, and we find $\Phi_{bc} = 5.0\degr^{\,+3.9\degr}_{\,-2.3\degr}$. 

\subsubsection{Secular behavior}
The orbital parameters we identified are only valid for the reference epoch (HJD\,2\,452\,490.0) because the system configuration is varying with time. Therefore, the mutual inclination angle we determined is only a snapshot of the system and could be misleading about its normal characteristics. For example, we might have caught the system when planets b's and c's orbits were near their minimum or maximum mutual inclination. 

\begin{figure*}[ht!]
\resizebox{\hsize}{!}{\includegraphics{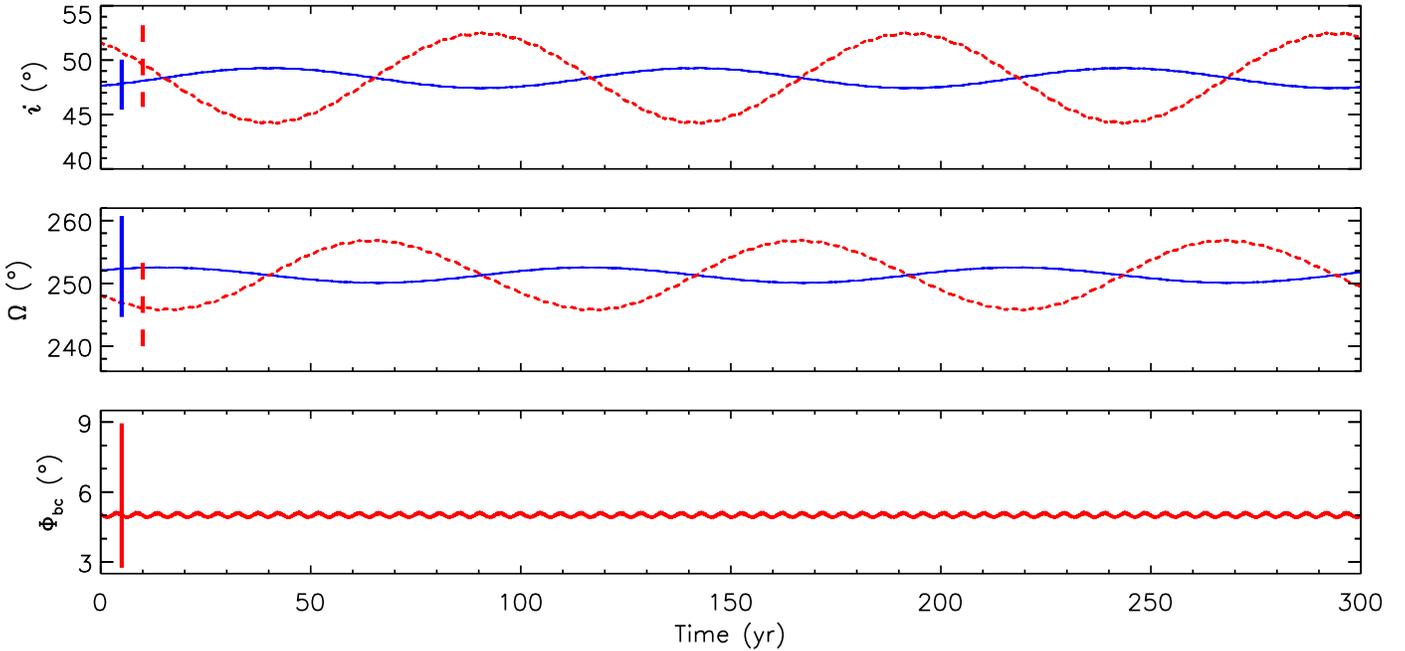}}
\caption{Long-term evolution of the projected orbital elements and mutual inclination for planets b (solid line) and c (dashed line). The variation is regular and continues in a similar fashion for at least 1\,Myr. The bars on the left indicate the uncertainties in the osculating orbital elements used to initialize the simulation.}
\label{f4}
\end{figure*}

To study the time-dependency of the configuration implied by our model for the GJ\,876 system, we integrated the planets' orbital motion forward for 1\,Myr. A short segment of the results for the inclinations, longitudes of the ascending nodes, and mutual inclination for the orbits of planets b and c are shown in Fig.~\ref{f4}. We find that the orbital projection angles for the planets are varying regularly, but with a variety of different frequencies and amplitudes. The mutual inclination between the planets' orbits varies with a main period of 4.8\,yr and amplitude 0.15$\degr$. There are also two other coherent lower-amplitude periodicities around 60\,d, which is similar to the outer planet's orbital period. These low-amplitude variations are slightly out of phase with one another and their interference leads to beat patterns over 10\,yr timescales.

Aside from mutual inclination changes, the two planets' orbital orientation angles relative to the plane of the sky librate with a period of 101.9\,yr. This is a projection effect arising from the libration of the planets' orbital nodes. Their mutual inclination is not affected by this variation and so the planets' orbital nodes must be librating together. We conclude from this exploratory investigation that our measured mutual inclination for the orbits of planets b and c is likely representative of the long-term status of the system as the measurement uncertainties are an order of magnitude larger than the potential variations. A more thorough examination of the dynamical qualities of our model would be interesting, but is beyond the scope of the current paper.

\section{Discussion}
Our analysis has revealed the full three-dimensional orbits, and thus the degree of coplanarity, of two planets in an exoplanetary system around a normal star for the first time\footnote{A previous coplanarity measurement was obtained for two planets orbiting a pulsar \citep{konacki03}.}. Broadly speaking, we find the orbits of GJ\,876b and c to be coplanar. However, our results also imply a small, but potentially significant ($\sim$95\% confidence), non-zero orbital mutual inclination that could be important.

Dynamical friction from planetesimals during the late stages of planet formation is expected to result in orbits for gas giants that are coplanar \citep{kokubo95,pollack96,goldreich04}. Therefore, our general result provides further evidence for planet formation in a circumstellar disk, and suggests that the evolution of planetary systems might not lead to excitation of extreme inclinations for planetary orbits relative to the original plane of the disk. Additional evidence from observations of exoplanetary systems for planet formation in a disk and little inclination from the original plane includes the coplanarity of an exoplanet's orbit with a debris disk \citep{benedict06} and the stellar spin -- planet orbit alignment of a number of transiting planet systems \citep{winn05, winn06, wolf07, winn07, narita07, bouchy08, winn08, loeillet08, johnson08, cochran08}, although see \citet{hebrard08} for one possible exception to this trend.

Our more subtle finding of a possible small degree of non-coplanarity in the GJ\,876 system is a complement to the observations that planetary system evolution often leads to eccentricity excitation and displacement of planets from their birthplace. As has been previously noted, the eccentricity of planet c is significantly non-zero (0.27), and it is unlikely that both planets b and c could have formed in situ \citep{laughlin05}. Therefore, it seems probable that the system has undergone some significant evolutionary changes.

\citet{lee02} have suggested that convergent migration of planets b and c due to disk torques led to resonance capture and eccentricity excitation. This seems to be the most likely explanation for the system's configuration, but there is still an open question of how the planets' eccentricities were kept from being excited to even higher values while the planets were migrating \citep{kley04, kley05, laughlin05}. Either the planets actually didn't migrate very far, or there was effective eccentricity damping during the migration. 

Along this same line, \citet{thommes03} have shown that resonance capture of two planets can also result in an inclination-type mean motion resonance that quickly leads to excitation of mutual inclinations of 30$\degr$. Mutual inclinations of 60$\degr$ or more can be achieved if the system experiences this simultaneously with an eccentricity-type mean motion resonance. However, entry into the inclination-type mean motion resonance requires the eccentricity of the inner planet to be $\goa$0.6, which is a condition that was likely not met in the GJ\,876 system. Our finding of only a small mutual inclination for planets b and c is therefore a further constraint on the system's evolutionary history. The nearly coplanar configuration of the planets' orbits is fully consistent with the scenario that they did not experience the inclination-type mean motion resonance because of the only moderate eccentricity excitation during migration. Thus, the question of why the planets' eccentricities were not excited to higher values becomes more important. It would be interesting to investigate whether hydrodynamic simulations of differential migration and resonance capture due to disk interactions \citep[e.g.][]{kley04} could reproduce the small degree of non-coplanarity we have found when extended to three dimensions. 

As the GJ\,876 system is the only planetary system other than the Solar System for which we have tight constraints on the degree of coplanarity it is interesting to compare the two. Surprisingly, we find that they share some similarities despite their obvious differences. GJ\,876b and c have a mass ratio ($m_{b}/m_{c}$ = 3.38) very similar to the Jupiter -- Saturn pair ($m_{Jup}/m_{Sat}$ = 3.34). This seems like a coincidence because it is unclear how such a property of neighboring gas giants could be maintained in different formation environments. Gas giants are thought to form via runaway gas accretion on to a solid core so such a property would require exact timing uniquely for each case.

More interestingly, \citet{tsiganis05} hypothesized that Jupiter and Saturn experienced an encounter with the 2:1 mean motion resonance due to migration. They suggested that this encounter led to eccentricity and mutual inclination excitation for the planets in the outer part of the Solar System, and is the reason for their currently non-circular and non-coplanar orbits. In contrast to the GJ\,876 system though, the \citet{tsiganis05} model for the Solar System has Jupiter and Saturn passing through the resonance owing to their diverging migration. As a result of being caught in the resonance, GJ\,876c's eccentricity was pumped up to at least three times the value that any of the Solar System giant planets' orbits reach. Additionally, our results indicate that the GJ\,876 b-c orbital mutual inclination is potentially a few times larger as well. Furthermore, the Jupiter -- Saturn 2:1 resonance encounter interactions also involved Uranus and Neptune, and the planets' final orbits depended on the details of the complex four body scattering. The GJ\,876 system is known to harbor an additional low-mass planet in a short period orbit. It is unclear how this object was involved in the dynamical evolution of the system, to say nothing of other still-to-be-discovered planets that could potentially exist in the system.

Ultimately, our determination of the orbital mutual inclination for GJ\,876b and c is just one more piece of the planet formation and evolution puzzle. It would be useful to obtain such measurements for many other exoplanetary systems to see if most or all systems tend to be fairly coplanar, but with a small amount of mutual inclination. Systems with planets in low-order resonances are particularly interesting targets due to the constraints on disk interactions knowledge of their architecture provides. In this regard, our analysis illustrates how such measurements could be achieved using data obtained from existing facilities. The GJ\,876 system is unique for the size and timescale of the planet-planet interactions, and so radial velocity data for other systems will not be as sensitive to their architecture. However, a number of other moderately interacting multi-planet systems are better astrometric targets than GJ\,876 because one of the planets in them induces a host star perturbation larger than 1\,mas, which is the typical measurement uncertainty of the \textit{HST} FGS. If robust astrometric characterization of one planet in a moderately interacting system can be obtained, then dynamical considerations could be used to constrain the degree of coplanarity for the other planets.

\begin{acknowledgements}
We thank Ansgar Reiners and an anonymous referee for comments that helped us improve this paper. J.B. and A.S. acknowledge funding for this work from the DFG through grants GRK 1351 and RE 1664/4-1.
\end{acknowledgements}

\bibliographystyle{aa}
\bibliography{ms}

\end{document}